\documentclass[runningheads]{llncs}
\usepackage{amssymb}
\usepackage{amsmath}
\usepackage{graphicx}
\usepackage{color}
\usepackage[algo2e,ruled,vlined]{algorithm2e}

\begin{document}

\pagestyle{headings}
\mainmatter

\title{Complexity and Computation of Connected \\Zero Forcing}
\titlerunning{On Connected Zero Forcing}

\author{Boris Brimkov}
\authorrunning{B. Brimkov}
\institute{
Department of Computational \& Applied Mathematics,\\ Rice University, Houston, TX 77005, USA\\
\email{boris.brimkov@rice.edu}
}

\maketitle

\begin{abstract}

Zero forcing is an iterative graph coloring process whereby a colored vertex with a single uncolored neighbor forces that neighbor to be colored. 
It is NP-hard to find a minimum zero forcing set -- a smallest set of initially colored vertices which forces the entire graph to be colored. We show that the problem remains NP-hard when the initially colored set induces a connected subgraph. We also give structural results about the connected zero forcing sets of a graph related to the graph's density, separating sets, and certain induced subgraphs, and we characterize the cardinality of the minimum connected zero forcing sets of unicyclic graphs and variants of cactus and block graphs. Finally, we identify several families of graphs whose connected zero forcing sets define greedoids and matroids.

\smallskip


\end{abstract}

\section{Introduction}

Zero forcing is an iterative graph coloring process where at each time step, a colored vertex with a single uncolored neighbor forces that neighbor to be colored; the zero forcing number of a graph is the cardinality of the smallest set of initially colored vertices which causes the entire graph to be colored. Zero forcing was introduced in an AIM workshop on linear algebra and graph theory in 2006 \cite{AIM-Workshop} and was used to bound the maximum nullity (equivalently, the minimum rank) of the family of symmetric matrices described by a graph. Despite being NP-hard to compute \cite{aazami}, the zero forcing number is generally more attainable than the maximum nullity, which makes it a valuable tool in the study of this algebraic parameter. In addition to its original linear algebraic application, zero forcing has found a variety of uses in physics, logic circuits, coding theory, power network monitoring, and in modeling the spread of diseases and information in social networks; see \cite{quantum1,logic1,powerdom3,powerdom2} for more details. The zero forcing number has also been used to bound or approximate various other graph parameters \cite{zf_tw,zf_np}. Closed formulas, characterizations, and bounds for the zero forcing number have been derived for graphs with special structure (cf. \cite{AIM-Workshop,benson,Eroh,Meyer}). 

A natural graph theoretic variant of zero forcing is obtained by requiring every set of initially colored vertices to induce a connected subgraph. This extension of zero forcing, called connected zero forcing, was introduced by Brimkov and Davila in \cite{brimkov}; that paper explored the differences and similarities between zero forcing and connected zero forcing, established several structural results about connected zero forcing sets, and characterized the connected zero forcing numbers of several families of graphs. A concurrent paper by Davila et al. \cite{CF_paper} explored bounds on the connected zero forcing number in terms of other graph parameters.

Studying connected zero forcing can further the understanding of the forcing process and the underlying structure of forcing sets in general. Moreover, in a connected graph $G$, the connected zero forcing number is a sharp upper bound to the maximum nullity, path cover number, chromatic number minus one, and power domination number of $G$ \cite{brimkov,CF_paper}. Requiring a zero forcing set to be connected also has meaningful interpretations in many of the physical phenomena modeled by zero forcing. For example, it is often the case that ideas or diseases originate from a single connected source in a social network or geographic region; thus, connected zero forcing may be better suited to model propagation in those scenarios. As another example, in the application of zero forcing to power network monitoring,\footnote{See \cite{benson} for a more thorough introduction to the power domination problem and its connection to zero forcing.} one could imagine a scenario where in addition to the production cost of the phase measurement units, there is a significant cost to dispatch a technician to install and maintain the units. Thus, an electric power company may seek to place all measurement devices in a compact, connected region in the network so that a technician can be sent on a single trip to install or maintain the devices, in addition to installing the smallest number of devices necessary to monitor the entire system.

Other variants of zero forcing, such as positive semidefinite zero forcing and signed zero forcing, have also been studied. These are typically obtained by modifying the color change rule or adding certain restrictions to zero forcing, and are often designed to bound different linear algebraic parameters. For example, in the positive semidefinite variant, the zero forcing color change rule acts separately on certain induced subgraphs; the minimum cardinality sets which force a graph using this modified rule are used to study the maximum nullity of the positive semidefinite matrices described by the graph (cf. \cite{Barioli,positive_zf2}). Similarly, the signed variant can be used to bound the maximum nullity of a matrix with a given sign pattern \cite{signed_zf}.

Another related graph parameter whose connected variant has been investigated is the domination number --- the minimum cardinality of a vertex set $S$ which contains or is adjacent to every vertex in the graph; requiring $S$ to be connected results in the connected domination number, which has distinct properties and applications. Connected domination has been extensively studied, e.g., in \cite{Sampathkumar,Caro,Desormeaux}; both domination and connected domination are NP-complete \cite{GJ}, with the latter generally being harder to solve exactly. Nevertheless, there are some strategies to improve on a brute force enumeration algorithm despite the non-locality of the connected domination problem (see, e.g., \cite{connected_dom_1}). 

As one of the main results of this paper, we establish the NP-completeness of connected zero forcing. Thus, as with zero forcing, this problem cannot be solved efficiently in general, but there can be bounds in terms of other graph parameters and characterizations for specific graphs. To this end, we give two lower bounds on the connected zero forcing number in terms of certain vertices and induced subgraphs in the graph. We also characterize the connected zero forcing numbers of unicyclic graphs and some variants of cactus and block graphs. Related parameters of such graphs have been investigated in the past (and sometimes rediscovered): for example, \cite{pathcover1} and \cite{pathcover2} give polynomial time algorithms for the path cover number of trees; \cite{johnson} and \cite{AIM-Workshop} respectively show that for trees, the path cover number equals the maximum nullity and the zero forcing number; \cite{brimkov} gives a linear time algorithm for the connected zero forcing number of trees and characterizes graphs in which it equals the zero forcing number; \cite{AIM-Workshop} and \cite{Huang} show that in block graphs, the maximum nullity equals the zero forcing number; \cite{fallat} surveys several characterizations and gives polynomial time algorithms for the maximum nullity and path cover number of unicyclic graphs; \cite{row} and \cite{taklimi} show that the zero forcing number of unicyclic graphs and cactus graphs equals the path cover number; \cite{power_dom_block} characterizes the power domination number of block graphs.

In general, since connectivity is a global property, non-local problems like connected zero forcing are typically much harder to solve exactly than their non-connected analogues. However, there are some other simple cases where efficient computation is possible. If the connected zero forcing number $Z_c(G)$ is known to be very small or very large, an enumeration approach can be used to find a minimum connected zero forcing set in polynomial time. For example, if $k_1\leq Z_c(G)\leq k_2<\frac{n}{2}$, it can be checked whether each of the $\binom{n}{k_1}+\cdots+\binom{n}{k_2}$ sets of vertices of appropriate size is connected and forcing in $O(n^2)$ time, so $Z_c(G)$ can be computed in $O((k_2-k_1)n^{2+k_2})$ time. An enumeration approach can also be used to efficiently compute the connected zero forcing number of graphs with polynomially many connected induced subgraphs; see \cite{meeks} for another dynamic graph coloring process which can be solved efficiently in such graphs. In the last part of the paper, we identify some graphs in which even a greedy algorithm can be used to obtain a minimum connected zero forcing set; for some of these graphs, the collection of all connected zero forcing sets can be used to define greedoids and matroids. 

The paper is organized as follows. In the next section, we recall some graph theoretic notions, specifically those related to zero forcing. In Section 3, we obtain some novel structural results about connected zero forcing, and recall some results from \cite{brimkov} which are used in the sequel. In Section 4, we prove that connected zero forcing is NP-complete. In Section 5, we give closed formulas for the connected zero forcing numbers of unicyclic graphs, and variants of cactus graphs and block graphs. In Section 6, we establish a connection between connected zero forcing sets and accessible set systems. We conclude with some final remarks and open questions in Section 7.

\section{Preliminaries}

\subsection{Graph theoretic notions}
A graph $G=(V,E)$ consists of a vertex set $V$ and an edge set $E$ of two-element subsets of $V$. The \emph{order} and \emph{size} of $G$ are denoted by $n=|V|$ and $m=|E|$, respectively. Two vertices $v,w\in V$ are \emph{adjacent}, or \emph{neighbors}, if $\{v,w\}\in E$. If $v$ is adjacent to $w$, we write $v\sim w$; otherwise, we write $v\not\sim w$. The \emph{neighborhood} of $v\in V$ is the set of all vertices which are adjacent to $v$, denoted $N(v;G)$; the dependence on $G$ can be omitted when it is clear from the context. The \emph{degree} of $v\in V$ is defined as $d(v;G)=|N(v;G)|$. The minimum degree and maximum degree of $G$ are denoted by $\delta(G)$ and $\Delta(G)$, respectively. Given $S \subset V$, the \emph{induced subgraph} $G[S]$ is the subgraph of $G$ whose vertex set is $S$ and whose edge set consists of all edges of $G$ which have both endpoints in $S$. The number of connected components of $G$ will be denoted by $\kappa(G)$, and an isomorphism between graphs $G_1$ and $G_2$ will be denoted by $G_1\simeq G_2$.

A \emph{leaf}, or \emph{pendant}, is a vertex with degree 1. 
An \emph{articulation point} (also called a \emph{cut vertex}) is a vertex which, when removed, increases the number of connected components in $G$. Similarly, a \emph{bridge} (also called a \emph{cut edge}) is an edge which, when removed, increases the number of components of $G$. A \emph{biconnected component}, or \emph{block}, of $G$ is a maximal subgraph of $G$ which has no articulation points. An \emph{outer block} is a block with at most one articulation point. 
A \emph{unicyclic graph} is a graph with exactly one cycle. A \emph{cactus graph} is a graph in which every block is a cycle or a cut edge, and a \emph{block graph} is a graph in which every block is a clique. For other graph theoretic terminology and definitions, we refer the reader to \cite{bondy}.

\subsection{Zero forcing}
Given a graph $G=(V,E)$ and a set $S \subset V$ of initially colored vertices, the \emph{color change rule} dictates that at each integer-valued time step, a colored vertex $u$ with a single uncolored neighbor $v$ \emph{forces} that neighbor to become colored; such a \emph{force} is denoted $u\rightarrow v$. 
The \emph{derived set} of $S$ is the set of colored vertices obtained after the color change rule is applied until no new vertex can be forced; it can be shown that the derived set of $S$ is uniquely determined by $S$ \cite{AIM-Workshop}. A \emph{zero forcing set} is a set whose derived set is all of $V$; the \emph{zero forcing number} of $G$, denoted $Z(G)$, is the minimum cardinality of a zero forcing set. 

A \emph{chronological list of forces} of $S$ is a sequence of forces applied to obtain the derived set of $S$ in the order they are applied; there can also be initially colored vertices which do not force any vertex. 
Generally, the chronological list of forces is not uniquely determined by $S$; for example, it may be possible for several colored vertices to force an uncolored vertex at a given step. 
A \emph{forcing chain} for a chronological list of forces is a maximal sequence of vertices $(v_1,\ldots,v_k)$ such that $v_i\rightarrow v_{i+1}$ for $1\leq i\leq k-1$. A \emph{singleton} chain is a forcing chain consisting of a single vertex, i.e., an initially colored vertex which does not force any vertex. If a vertex forces another vertex at some step of the forcing process, then it cannot force a second vertex at a later step, since that would imply it had two uncolored neighbors when it forced for the first time. Thus, each forcing chain induces a distinct path in $G$, one of whose endpoints is an initially colored vertex, and all other vertices are uncolored at the initial time step; we will say the initially colored vertex \emph{initiates} the forcing chain. The set of all forcing chains for a chronological list of forces is called the \emph{chain set}, and is uniquely determined by the chronological list of forces. Any chain set of a zero forcing set forms a path cover of $G$.

A \emph{connected zero forcing set} of $G$ is a zero forcing set of $G$ which induces a connected subgraph. The \emph{connected zero forcing number} of $G$, denoted $Z_c(G)$, is the cardinality of a minimum connected zero forcing set of $G$. For short, we may refer to these as \emph{connected forcing set} and \emph{connected forcing number}. Note that a disconnected graph can never have a connected forcing set.

\section{Structural results and technical lemmas}

An important concept to studying and understanding the zero forcing process is that of \emph{zero forcing spread} of a vertex $v$ and edge $e$ in graph $G$; these parameters, defined as $z(G;v)=Z(G)-Z(G-v)$ and $z(G;e)=Z(G)-Z(G-e)$, respectively, describe the effects of deleting a vertex or edge from the graph on the zero forcing number of the graph. It has been shown in \cite{Edholm,Huang} that the zero forcing spread of any vertex or edge is bounded by 1; more precisely, for any graph $G$, vertex $v$, and edge $e$, $-1\leq z(G;v)\leq 1$ and $-1\leq z(G;e)\leq 1$. In \cite{brimkov}, the analogous concept of \emph{connected forcing spread} of a non-articulation vertex $v$ was defined as $z_c(G;v)=Z_c(G)-Z_c(G-v)$, and it was shown that unlike the zero forcing spread, the connected forcing spread of a vertex can be arbitrarily large:

\begin{proposition}\emph{\cite{brimkov}}
For any $c_1<0$ and $c_2>0$, there exist graphs $G_1$ and $G_2$ and vertices $v_1\in G_1$ and $v_2\in G_2$ such that $z_c(G_1;v_1)<c_1$ and $z_c(G_2;v_2)>c_2$. 
\end{proposition}

We now show that the same is true of the connected forcing spread of an edge $e$, which we define as $z_c(G;e)=Z_c(G)-Z_c(G-e)$. In this definition, we restrict $e$ to be a non-cut edge of $G$, since a disconnected graph cannot have a connected forcing set. In particular, we show that unlike the zero forcing spread, the connected forcing spread of an edge can be arbitrarily large. 

\begin{proposition}
For any $c_1<0$ and $c_2>0$, there exist graphs $G_1$ and $G_2$ and edges $e_1\in G_1$ and $e_2\in G_2$ such that $z_c(G_1;e_1)<c_1$ and $z_c(G_2;e_2)>c_2$. 
\end{proposition}

\begin{figure}[ht!]
\begin{center}
\includegraphics[scale=0.35]{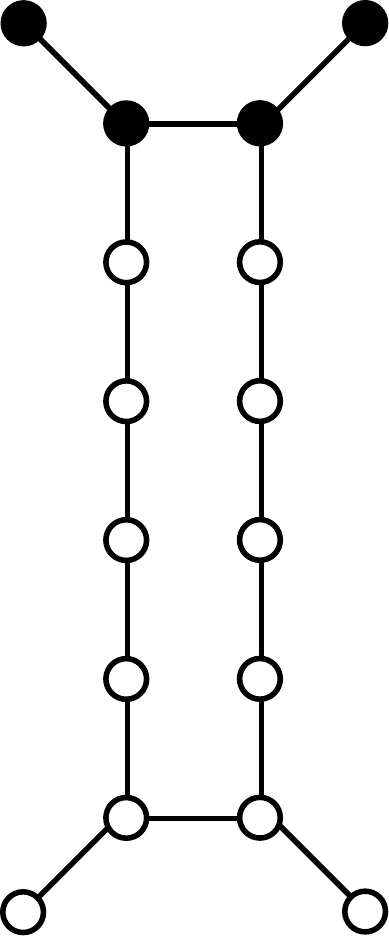}\qquad\qquad
\includegraphics[scale=0.35]{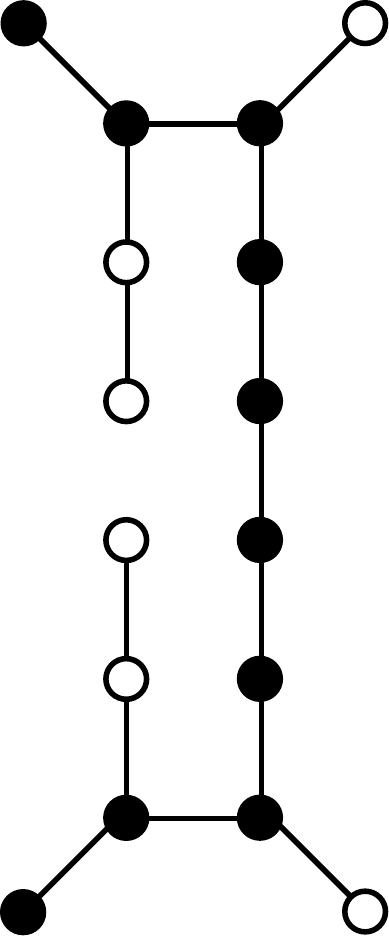}\qquad\qquad\qquad\qquad\qquad\qquad
\includegraphics[scale=0.35]{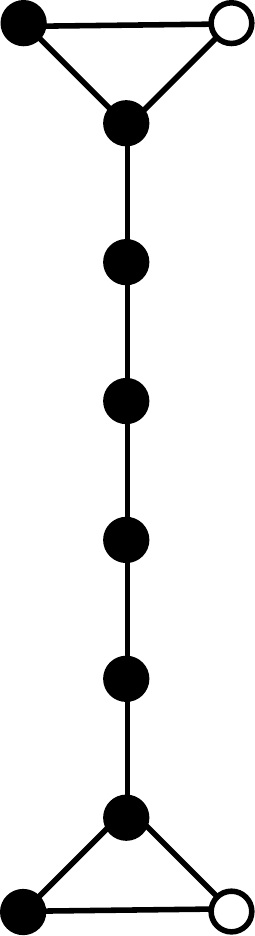}\qquad\qquad
\includegraphics[scale=0.35]{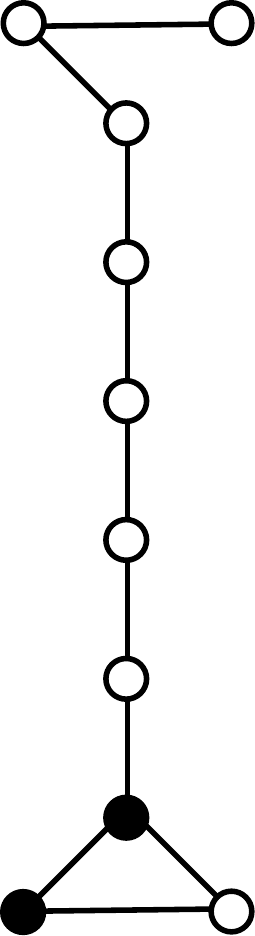}
\caption{\emph{Left:} Deleting an edge from $G_1$ makes $Z_c(G_1)$ increase arbitrarily. \emph{Right:} Deleting an edge from $G_2$ makes $Z_c(G_2)$ decrease arbitrarily.}
\label{spread}
\end{center}
\end{figure}

\proof
Let $G_1$ be the graph obtained by appending a pendant vertex to each endpoint of two maximally distant edges of an even cycle $C_{2k}$, $k\geq 4$, and let $e$ be an edge neither of whose endpoints are adjacent to a pendant; see Figure \ref{spread}, left for an illustration. It is easy to see that $Z_c(G_1)=4$ and $Z_c(G_1-e)=k+4$. Thus, $z_c(G_1;e)=-k$, which can be made smaller than any constant $c_1$.

Let $G_2$ be the graph obtained by appending a copy of $K_3$ to each end of a path $P_k$, and let $e$ be an edge whose endpoints have degrees 2 and 3; see Figure \ref{spread}, right for an illustration. It is easy to see that $Z_c(G_2)=k+2$ and $Z_c(G_2-e)=2$. Thus, $z_c(G_2;e)=k$, which can be made larger than any constant $c_2$.
\qed
\vspace{9pt}
Another important matter to consider when studying the forcing process is the relationship between the density of a graph and its connected forcing number. It can be readily verified that sparse graphs can have both large and small zero forcing numbers and connected forcing numbers; path graphs and star graphs are extremal in this regard. The following theorem shows that in contrast, dense graphs can only have ``large'' connected forcing numbers and zero forcing numbers. 

\begin{theorem}
\label{density}
Let $G=(V,E)$ be a graph with $|E|=\Omega(|V|^2)$. Then, $Z_c(G)=\Theta(|V|)$.
\end{theorem}
\proof
Let $n=|V|$ and suppose for contradiction that for every $S\subset V$, $\delta(G[S])=o(n)$. Let $G_0=G$; for $1\leq i\leq n$, let $v_i$ be a vertex such that $d(v_i;G_{i-1})=\delta(G_{i-1})$ and let $G_i=G_{i-1}-v_i$. In words, the graphs $\{G_i\}_{i=1}^n$ are obtained by repeatedly deleting a vertex of minimum degree. By our assumption, for $1\leq i\leq n$, $\delta(G_i)=o(n)$ so each $G_i$ has $o(n)$ fewer edges than $G_{i-1}$. However, this is a contradiction, since $n\cdot o(n)\neq \Omega(n^2)$. Thus, there must be some $S\subset V$ for which $\delta(G[S])=\Omega(n)$. 

Let $R$ be a minimum connected forcing set of $G$; clearly $|R|=O(n)$. Fix some chronological list of forces, and let $v$ be the first vertex in $S$ (if any) which forces another vertex. At that step of the forcing process, $v$ and all-but-one of its neighbors must be colored; since $\delta(G[S])=\Omega(n)$, there must be $\Omega(n)$ colored vertices at the step when $v$ performs a force. Since $v$ is the first vertex in $S$ to perform a force, each of $v$'s neighbors in $S$ is either in $R$, or has been forced by a distinct forcing chain (since if two vertices in $S$ are in the same forcing chain, the one that comes first in the chain would have performed a force before $v$). If no vertex of $S$ ever performs a force, then again each vertex in $S$ is either in $R$, or has been forced by a distinct forcing chain. Since each forcing chain is initiated by a unique element in $R$, and since $|R|\geq |S|\geq \delta(G[S])=\Omega(n)$, it follows that $Z_c(G)=\Theta(n)$. 
\qed
\vspace{9pt}
A similar argument as above can be used to show that for $G=(V,E)$ with $|E|=\Omega(|V|^2)$, then $Z(G)=\Theta(|V|)$, as well. It should be noted that Theorem~\ref{density} describes only the asymptotic relationship between the density of a graph and its connected forcing number. It is also useful to obtain non-asymptotic bounds on the connected forcing number in terms of the edge count and other easily computable parameters; some progress to this end has been made in \cite{CF_paper}. The next results in this section are also a step in this direction. 

The following lemma generalizes a result from \cite{brimkov} regarding vertices which belong to every connected forcing set. In contrast, it has been shown that no vertex belongs to every zero forcing set \cite{Barioli}. 

\begin{lemma}
\label{main_lemma}
Let $G$ be a connected graph, $S$ be a separating vertex set of $G$, and $V_1,\ldots,V_k$ be the vertex sets of the connected components of $G-S$. If each vertex of $S$ is incident to each $V_i$, $1\leq i\leq k$, then every connected forcing set of $G$ contains a vertex from at least $k-1$ of $V_1,\ldots,V_k$. Moreover, if $k=2$ and $Z(G[V_i])> |S|$ for $i\in\{1,2\}$, or if $k\geq 3$, then every connected forcing set of $G$ contains a vertex of $S$.
\end{lemma}

\proof
Let $R$ be an arbitrary connected forcing set of $G$ with an arbitrary chronological list of forces, and suppose $R$ does not contain vertices from two components of $G-S$, say $G[V_1]$ and $G[V_2]$. Let $v$ be the first vertex in $V_1\cup V_2$ to be forced; since $N(v;G)\subset S\cup V_1\cup V_2$, $v$ must be forced by a vertex of $S$. However, no vertex of $S$ can force $v$, since at that step every vertex of $S$ has at least two uncolored neighbors --- one in $V_1$ and one in $V_2$. Thus, $R$ must contain a vertex from at least $k-1$ of $V_1,\ldots,V_k$. In particular, if $k\geq 3$, $R$ must contain a vertex from at least two components of $G-S$. Since $R$ is connected, and since any path between two vertices from different components of $G-S$ must contain a vertex of $S$, $R$ contains a vertex of $S$. 

Now suppose $k=2$, and suppose for contradiction that $R\subset V_1$. Let $Z$ be the set of vertices of $V_2$ forced by vertices in $S$. We claim that $Z$ is a zero forcing set of $G[V_2]$ and that the list of forces where the $i^\text{th}$ force is the $i^\text{th}$ instance of a vertex of $V_2$ forcing another vertex of $V_2$ in the chronological list of forces of $R$ in $G$, is a chronological list of forces for $Z$ in $G[V_2]$.
To see why, note that if $v\in V_2$ forces another vertex of $V_2$ at some step of the forcing process of $G$, by induction and since $N(v;G[V_2])\subset N(v;G)$, $v$ and all-but-one of its neighbors are colored in $G[V_2]$ at the corresponding step of the forcing process of $G[V_2]$. Thus, $v$ would be able to force the same vertex in $G[V_2]$ as in $G$, so each force between two vertices of $V_2$ in $G$ can also be performed in $G[V_2]$. Since in $G$, each vertex in $V_2$ is forced either by a vertex of $S$ or a vertex of $V_2$, in $G[V_2]$ each vertex is either in $Z$ or is in a forcing chain initiated by a vertex in $Z$; thus $Z$ is a forcing set of $G[V_2]$.
However, $|Z|\leq |S|$ since each vertex in $S$ forces at most one vertex of $V_2$ in $G$; this contradicts the assumption that $Z(G[V_2])>|S|$. Thus, $R\not\subset V_1$; similarly, $R\not\subset V_2$, and $R\not\subset V_1\cup V_2$, since $R$ is connected and $G[V_1\cup V_2]$ is not. Thus, $R$ contains a vertex of $S$.
\qed
\vspace{9pt}
\noindent We now fix some terminology and notation which will be used in the sequel.


\begin{definition}
A \emph{pendant path attached to vertex} $v$ in graph $G=(V,E)$ is a set $P\subset V$ such that $G[P]$ is a path component of $G-v$, one of whose ends is adjacent to $v$ in $G$. The neighbor of $v$ in $P$ will be called the \emph{base} of the path,
and $p(v)$ will denote the number of pendant paths attached to $v\in V$. 


\end{definition}

\begin{definition}
Let $G=(V,E)$ be a connected graph. Define 
\begin{eqnarray*}
R_1(G)&=&\{v\in V: \kappa(G-v)=2, \; p(v)=1\}\\
R_2(G)&=&\{v\in V: \kappa(G-v)=2, \; p(v)=0\}\\
R_3(G)&=&\{v\in V: \kappa(G-v)\geq 3\}\\
\mathcal{L}(G)&=&\bigcup_{v\in V}\{\text{all-but-one bases of pendant paths attached to } v\}\\
M(G)&=&R_2(G)\cup R_3(G)\cup \mathcal{L}(G). 
\end{eqnarray*} 
When there is no scope for confusion, the dependence on $G$ will be omitted.
\end{definition}

%

\begin{lemma}
\label{MR_lemma}
Let $G=(V,E)$ be a connected graph different from a path and $R$ be an arbitrary connected forcing set of $G$. Then $M\subset R$.
\end{lemma}
\proof
Since an articulation point $v$ is a separating set incident to each component of $G-v$, by Lemma \ref{main_lemma}, $R$ must contain $v$ for all $v\in R_2\cup R_3$. Moreover, if all components of $G-v$ are paths, then it is easy to verify that $R$ consists of $v$ and all-but-one bases of pendant paths attached to $v$, i.e. $\mathcal{L}\subset R$. Now, suppose $v$ is an articulation point such that not all components of $G-v$ are paths. If $R$ does not include any vertices from some component of $G-v$, that component must be a path, since otherwise the component cannot be forced by $v$ alone, or $R$ cannot be connected. Since by Lemma \ref{main_lemma}, $R$ includes a vertex from at least all-but-one components of $G-v$, and since the excluded component can only be a path, it follows that for each $u\in V$, $R$ includes at least all-but-one bases of pendant paths attached to $u$. By definition, $M=R_2\cup R_3\cup \mathcal{L}$, so $M\subset R$.
\qed

\vspace{9pt}
We recall a result from \cite{brimkov} which also concerns the set $M$ defined above. This result is used in characterizing the connected forcing numbers of other tree-like graphs in the following sections.

\begin{theorem}
\label{tree_thm}
\emph{\cite{brimkov}} Let $G$ be a tree different from a path; then $M$ is a minimum connected forcing set of $G$.
\end{theorem}

The next result is also related to vertices which belong to every connected forcing set.

\begin{proposition}
\label{block_prop}
Let $G$ be a connected graph different from a path and $B$ be a block of $G$ which is not a cut edge of a pendant path of $G$. Then every connected forcing set of $G$ contains at least $\delta(G[B])$ vertices of $B$.
\end{proposition}
\proof
Suppose there is a connected forcing set $S$ of $G$ which contains at most $\delta(G[B])-1$ vertices of $B$. Clearly there are uncolored vertices in $B$, since $B$ has at least $\delta(G[B])+1$ vertices. Any forcing chain initiated by a vertex outside $B$ and containing a vertex of $B$ must pass through an uncolored articulation point $p$ of $B$; by Lemma~\ref{MR_lemma}, $p\notin M$, so $p\in R_1$. However, this means $S$ contains a vertex of a pendant path, but not the vertex to which the path is attached --- this contradicts $S$ being connected or being forcing. Thus, there can be no forcing chain initiated outside $B$ and passing through $B$.

Now suppose there is a (non-singleton) forcing chain starting at $v\in B$ which contains another vertex of $B$. The vertex $v$ has at least $\delta(G[B])$ neighbors in $B$; however, by assumption, at most $\delta(G[B])-2$ of them can be colored. Since none of these neighbors of $v$ can get forced by a vertex outside of $B$, $v$ cannot force any vertex --- a contradiction. Thus, no uncolored vertex in $B$ can be forced, so $S$ must contain at least $\delta(G[B])$ vertices of $B$.
\qed
\vspace{9pt}
Using Proposition \ref{block_prop} and the fact that the only vertices which can belong to more than one block which is not part of a pendant path are the vertices in $R_2\cup R_3$, we formulate the following lower bound on the connected forcing number.

\begin{corollary}
\label{lower_bound}
Let $G$ be a connected graph, $\mathcal{B}$ be the set of blocks of $G$ which are not cut edges of pendant paths of $G$, and let $\mu(v)$ denote the number of blocks a vertex $v$ is part of. Then,
\begin{equation*}
Z_c(G) \geq \sum_{B\in \mathcal{B}}\delta(G[B])-\sum_{p\in R_2\cup R_3}(|\mu(p)|-1).
\end{equation*}
\end{corollary}

The bound in Corollary \ref{lower_bound} is tight, for example in a cycle or complete graph; this bound can be used in conjunction with the bound $Z_c(G)\geq |M|$ implied by Lemma \ref{MR_lemma}.

\section{NP-completeness of connected zero forcing}

In this section, we show that computing the connected forcing number of a graph is NP-complete. To begin, we state the decision version of this problem.\\

\noindent PROBLEM: Connected zero forcing ($CZF$)\\
INSTANCE: A simple undirected connected graph $G=(V,E)$ and a positive integer $k\leq |V|$.\\
QUESTION: Does $G$ contain a zero forcing set $S$ of size at most $k$ such that $G[S]$ is connected?

\begin{theorem}
\label{np_theorem}
$CZF$ is NP-complete.
\end{theorem}
\proof
We will first show that $CZF$ is in NP.
Given a set $S$ of vertices of $G$, it can be checked in polynomial time whether there is a vertex in $S$ with exactly one neighbor not in $S$. Moreover, there cannot be more than $|V|$ steps in a forcing process. Thus, a nondeterministic algorithm can check in polynomial time whether a subset of vertices of $V$ is forcing, whether it induces a connected subgraph, and whether it has size at most $k$. Thus, $CZF$ is in NP.

For our reduction, we select the problem of zero forcing, which was proved to be NP-complete in \cite{aazami}. The decision version of zero forcing is stated below. \\
PROBLEM: Zero forcing ($ZF$)\\
INSTANCE: A simple undirected graph $G=(V,E)$ and a positive integer $k\leq |V|$.\\
QUESTION: Does $G$ contain a zero forcing set $S$ of size at most $k$?

Next, we construct a transformation $f$ from $ZF$ to $CZF$.
Let $I=\langle G,k \rangle$ be an instance of $ZF$, where $G=(V,E)$ and $V=\{v_1,\ldots,v_n\}$. We define $f(I)=\langle G',k+2\rangle$, where $G'=(V\cup\{v^*,\ell_1,\ell_2\},E\cup\{\{v^*,v_i\}:1\leq i\leq n\}\cup\{\{v^*,\ell_1\},\{v^*,\ell_2\}\})$. See Figure 1 for an illustration of $G$ and $G'$.

\begin{figure}[ht!]
\begin{center}
\includegraphics[scale=0.35]{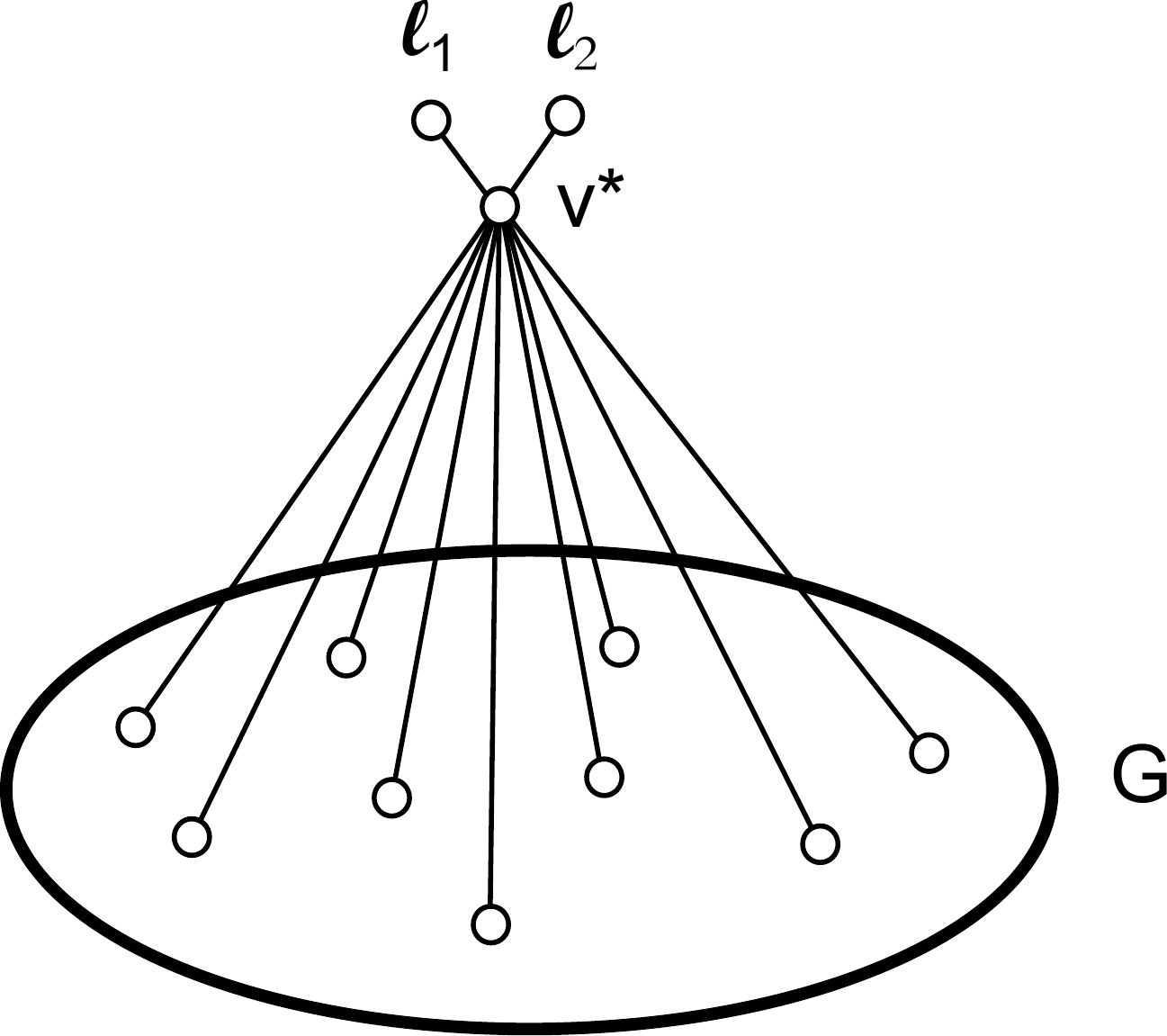}
\caption{Obtaining $G'$ from $G$.}
\label{fig1}
\end{center}
\end{figure}

Finally, we will prove the polynomiality and correctness of $f$. Clearly, $G'$ can be constructed from $G$ in polynomial time, so $f$ is a polynomial transformation. 

Suppose $I=\langle G,k\rangle$ is a \emph{`yes'} instance of $ZF$, i.e., that $G=(V,E)$ has a zero forcing set $S$ of size at most $k$. We claim that $S':=S \cup \{v^*,\ell_1\}$ is a connected forcing set of $G'$. To see why, first note that since $v^*$ is adjacent to every vertex in $S'-\{v^*\}$, $G'[S']$ is connected. Next, given an arbitrary chronological list of forces for $S$ in $G$, each force can also be applied for $S'$ in $G'$, since for any $v\in V$, $N(v;G')=N(v;G)\cup \{v^*\}$ and $v^*$ is initially colored; thus, when $v$ has a single uncolored neighbor in $G$ at some step of the forcing process, it will have the same uncolored neighbor in $G'$. When all vertices of $V$ in $G'$ are colored, $\ell_2$ will be the only uncolored vertex in $G'$, and it will be forced by $v^*$. Thus $S'$ is a connected forcing set of $G'$ of size at most $k+2$, so $f(I)=\langle G',k+2\rangle$ is a \emph{`yes'} instance of $CZF$.

Conversely, suppose $f(I)=\langle G',k+2\rangle$ is a \emph{`yes'} instance of $CZF$, i.e., that $G'$ has a connected forcing set $S'$ of size at most $k+2$. Fix a chronological list of forces for $S'$ in $G'$ and suppose $v^*$ forces a vertex $w\in V$. Then, both $\ell_1$ and $\ell_2$ must be in $S'$, since they are adjacent only to $v^*$, which cannot force them if it forces $w$. Moreover, $w$ must be the last uncolored vertex in $G'$, since if there was another uncolored vertex, $v^*$ would have more than one uncolored neighbor and could not force $w$. If $w$ is not an isolated vertex of $G$, then in the last step of the forcing process, $w$ can be forced by one of its neighbors in $V$ instead of by $v^*$. If $w$ is an isolated vertex of $G$, then it is a leaf of $G'$, and the set $S''=S'\backslash \{\ell_1\}\cup \{w\}$ is also a connected forcing set of $G'$, where $v^*$ does not force any vertex of $V$ (if we use the same chronological list of forces, except in the last step, $v^*\rightarrow \ell_1$ instead of $v^*\rightarrow w$). 

Thus, we can choose a connected forcing set $S'$ and a chronological list of forces for $S'$ such that $v^*$ does not force any vertex of $V$ in $G'$. We claim that $S:=S'\cap V$ is a forcing set of $G$. To see why, first note that $v^*$ must be in $S'$ by Lemma \ref{MR_lemma}, and that for any $v\in V$, $N(v;G)=N(v;G')\backslash \{v^*\}$. Thus, each force between vertices of $V$ in $G'$ can also be applied for $S$ in $G$, since if $v\in V$ has a single uncolored neighbor in $G'$ at some step of the forcing process, it will have the same uncolored neighbor in $G$. Moreover, since $v^*$ does not force any vertex in $V$, all vertices in $V$ must be forced by the elements of $S'$ which are in $V$. Thus, $S$ is a forcing set of $G$. Finally, to verify the size of $S$, note that by Lemma \ref{MR_lemma}, $v^*$ and at least one of $\ell_1$ and $\ell_2$ must be in $S'$, so $k+2\geq |S'|\geq |S'\cap V|+2=|S|+2$, so $S$ has size at most $k$. Thus, if $f(I)$ is a \emph{`yes'} instance of $CZF$, then $I$ is a \emph{`yes'} instance of $ZF$.
\qed

\vspace{9pt}
In view of Theorem \ref{np_theorem}, we cannot hope to efficiently compute the connected forcing number of an arbitrary graph. However, in the following sections, we investigate certain graphs whose connected forcing numbers can be found in linear time, and graphs whose connected forcing numbers can be found using a greedy algorithm.

\section{Characterizations of connected forcing numbers}

\subsection{Unicyclic graphs}
In this section, we will derive a closed formula for the connected forcing number of a unicyclic graph $G$ and give a linear time algorithm for finding a minimum connected forcing set of $G$. We first establish two technical lemmas which are applicable to arbitrary graphs that contain a cycle block.

Let $G$ be a connected graph and $C$ be the vertex set of a block of $G$ such that $G[C]$ is a cycle.  
Given vertices $u$ and $v$ of $C$, let $(u\hookrightarrow v)$ be the set of vertices of $C$ encountered while traveling counterclockwise from $u$ to $v$, not including $u$ and $v$; note that $(u\hookrightarrow u)$ is also well-defined. Let $(u\hookrightarrow)$ be the neighbor of $u$ which is counterclockwise of $u$ in $C$, and $(\hookleftarrow u)$ be the neighbor of $u$ which is clockwise of $u$ in $C$. We will refer to $(u\hookrightarrow v)$ as a \emph{segment} of $C$, and call $u$ and $v$ the \emph{ends} of the segment.

\begin{lemma}
\label{one_seg_lemma}
Let $G$ be a connected graph and $C$ be the vertex set of a block of $G$ such that $G[C]$ is a cycle. Then, any connected forcing set of $G$ can exclude at most one segment of $C$.
\end{lemma}
\proof
Let $R$ be an arbitrary connected forcing set of $G$. Suppose $(u\hookrightarrow v)$ and $(x\hookrightarrow y)$ are two non-intersecting and non-adjacent segments of $C$ which are not contained in $R$ (note that two intersecting or adjacent segments can be represented as a single segment). Without loss of generality, suppose $u$, $v$, $x$, and $y$ lie on $C$ in this counterclockwise order. Then $R$ contains at least one vertex between $v$ and $x$, and at least one vertex between $y$ and $u$; however, these vertices cannot be connected in $G[R]$ since all paths between them pass through the missing segments in $C$. Thus, there can be at most one segment of $C$ which is not contained in $R$.
\qed

\begin{lemma}
\label{two_ap_lemma}
Let $G$ be a connected graph and $C$ be the vertex set of a block of $G$ such that $G[C]$ is a cycle. A segment of $C$ excluded from a connected forcing set of $G$ can contain at most two articulation points, each of which is in $R_1(G)$.
\end{lemma}
\proof
Let $R$ be an arbitrary connected forcing set of $G$ and $(u\hookrightarrow v)$ be a segment of $C$ not contained in $R$; by Lemma \ref{MR_lemma}, $M\subset R$, so $(u\hookrightarrow v)$ cannot contain a vertex of $M$. Thus, each vertex in $(u\hookrightarrow v)$ is either a non-articulation point, or an articulation point in $R_1$; in the latter case, the entire pendant path attached to the vertex is also not in $R$ since otherwise $R$ could not be connected. Suppose $(u\hookrightarrow v)$ contains three distinct articulation points, $p$, $q$, and $r$, lying on $C$ in this counterclockwise order. Every path from a vertex of $C$ outside $(u\hookrightarrow v)$ to a vertex in $(p\hookrightarrow r)$ passes through $p$ or $r$. However, once $p$ and $r$ are forced by some forcing chains starting outside $(u\hookrightarrow v)$, each of $p$ and $r$ will have two uncolored neighbors and will not be able to force another vertex. Thus, the vertices in $(p\hookrightarrow r)$ cannot be forced; note that $(p\hookrightarrow r)\neq \emptyset$ since $q\in (p\hookrightarrow r)$. This contradicts $R$ being a forcing set, so $(u\hookrightarrow v)$ can contain at most two articulation points. 
\qed
\vspace{9pt}

\begin{lemma}
\label{rstarlemma}
Let $G$ be a unicyclic graph, $C$ be the vertex set of the cycle of $G$, and $(u^*\hookrightarrow v^*)$ be the largest segment of $C$ such that $R^*:=M\cup C\backslash (u^*\hookrightarrow v^*)$ is a forcing set of $G$. Then $R^*$ is a minimum connected forcing set of $G$.
\end{lemma}
\proof
The vertices in $V$ can be partitioned into $M$, $C\backslash M$, and $X$, where $X$ is the set of vertices in pendant paths of $G$ which are not in $M$; by Lemma \ref{two_ap_lemma}, 
$(u^*\hookrightarrow v^*)\subset C\backslash M$ and any articulation points in $(u^*\hookrightarrow v^*)$ are in $R_1$. Thus $V\backslash R^*=X\cup(u^*\hookrightarrow v^*)$, and deleting all vertices in $X\cup (u^*\hookrightarrow v^*)$ from $G$ does not disconnect it, so $R^*$ is a connected forcing set.

Now suppose there is a connected forcing set $R'$ of $G$ with $|R'|<|R^*|$. By Lemma \ref{MR_lemma}, $R'$ contains all vertices in $M$. Thus $R'$ must contain at most $|C\backslash M|-|(u^*\hookrightarrow v^*)|-1$ vertices of $(C\backslash M) \cup X$. By Lemma \ref{one_seg_lemma}, the vertices of $C$ not contained in $R'$ must form a segment $(u'\hookrightarrow v')$. If $R'=M\cup C\backslash (u'\hookrightarrow v')$, $(u'\hookrightarrow v')$ would be larger than $(u^*\hookrightarrow v^*)$, which contradicts our assumption about $(u^*\hookrightarrow v^*)$; thus, $R'$ includes some vertices of $X$. These vertices cannot be in pendant paths attached to vertices of $(u'\hookrightarrow v')$, since then $R'$ would be disconnected; if they are in pendant paths attached somewhere other than $u'$ and $v'$, then a set $R''$ without them is a smaller connected forcing set than $R'$, and we can henceforth consider $R''$ instead of $R'$. Similarly, if the vertices of $R'$ in $X$ are not the bases of the pendant paths containing them, then since $R'$ is connected, it must also include the bases of the pedant paths, and a set $R''$ without the non-base vertices of these pendant paths is a smaller connected forcing set than $R'$. Thus, without loss of generality, suppose the vertices of $R'$ in $X$ are the bases of pendant paths attached to $u'$ or $v'$. Then, $u'$ and $v'$ would be able to initiate forcing chains. Let $R''$ be obtained from $R'$ by replacing the vertices in $X$ by the vertices forced by $u'$ and $v'$. This resulting set is of the form  $M\cup C\backslash (u''\hookrightarrow v'')$, and has the same cardinality as $R'$, but $(u''\hookrightarrow v'')$ is larger than $(u^*\hookrightarrow v^*)$ --- a contradiction. Thus, no connected forcing set of $G$ can have cardinality less than $|R^*|$, so $R^*$ is a minimum connected forcing set of $G$.
\qed
\vspace{9pt}
In view of Lemma \ref{rstarlemma}, to find a minimum connected forcing set of a unicyclic graph $G$ with cycle $C$, one could generate all connected subgraphs of $C$, check whether each subgraph together with $M$ is forcing, and find the smallest one, in polynomial time. However, we will include a more thorough case analysis which reduces the number of segments that have to be compared, eliminates the need to check whether a set is forcing, and gives a linear time algorithm for finding a minimum connected forcing set of $G$.

\begin{figure}[ht!]
\begin{center}
\includegraphics[scale=0.20]{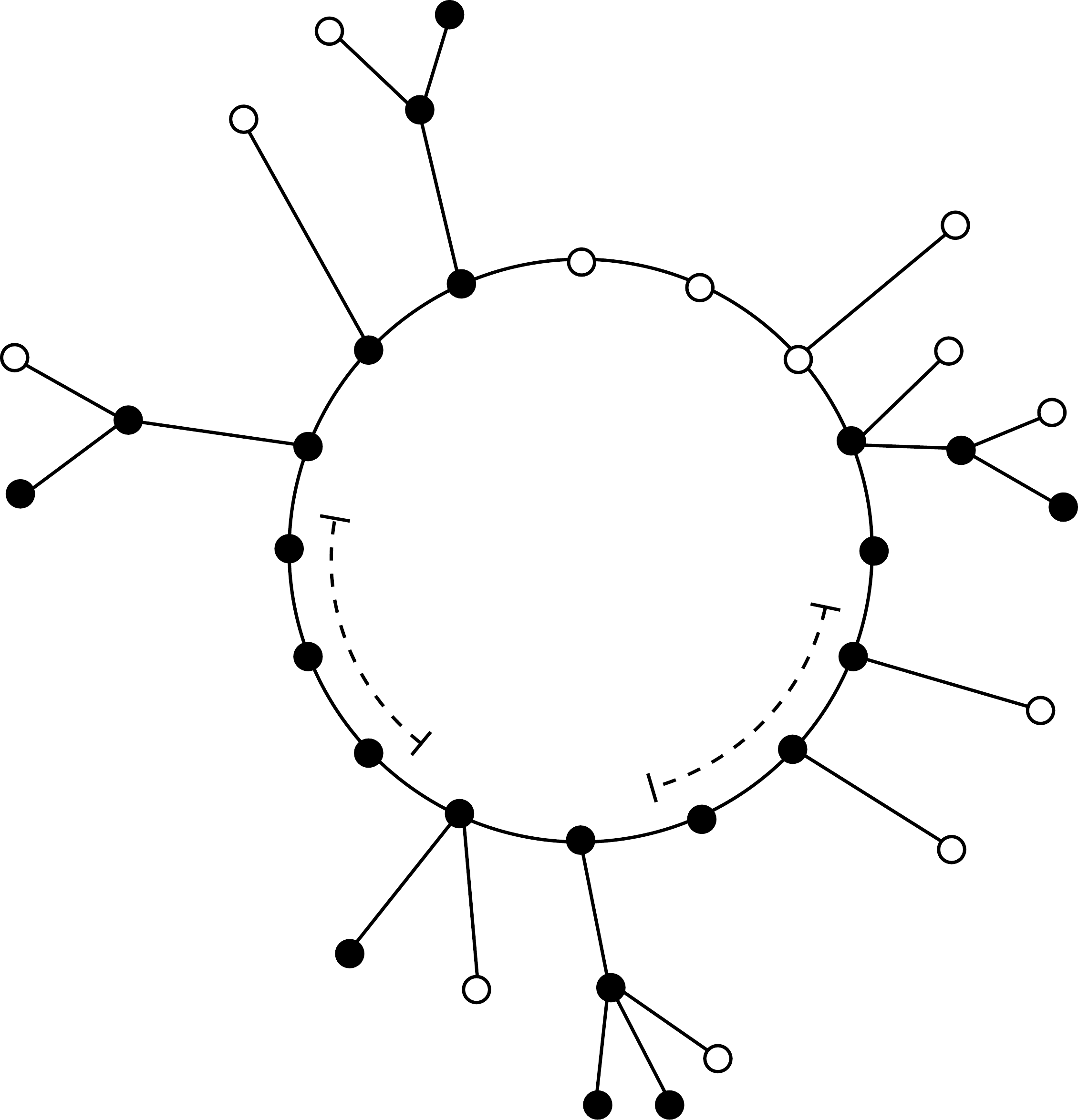}
\caption{A unicyclic graph and a minimum connected forcing set. Two other minimum connected forcing sets can be obtained by coloring the uncolored segment of $C$ and removing one of the segments indicated by the dashed lines.}
\label{fig_unicycle}
\end{center}
\end{figure}

To this end, we define a \emph{feasible} segment to be a segment $(u\hookrightarrow v)$ for which $R:=M\cup C\backslash(u\hookrightarrow v)$ is a forcing set of $G$ and which is maximal in this regard (with respect to inclusion). Clearly, $(u^*\hookrightarrow v^*)$ described in Lemma \ref{rstarlemma} is the largest feasible segment (or rather, \emph{a} largest feasible segment since there could be several feasible segments with the same maximum cardinality --- see, e.g., Figure \ref{fig_unicycle}). Let $A(C)=\{p_1,\ldots,p_k\}$ be the set of articulation points in $C$ in counterclockwise order. The following lemmas will allow us to enumerate the feasible segments of $C$; recall that $p(v)$ denotes the number of pendant paths attached to vertex $v$.

\begin{lemma}
\label{d2_lemma}
Let $G$ be a unicyclic graph, $C$ be the vertex set of the cycle of $G$ and suppose $|A(C)|\geq 3$. Let 
\begin{equation*}
f_2(u,v)=\begin{cases}
\{(u\hookrightarrow),(\hookleftarrow v)\} & \text{if }p(u)>0 \text{ and } p(v)>0\\
\{(u\hookrightarrow )\} & \text{if }(p(u)>0\text{ and } p(v)=0) \text{ or } ((p(u)=0 \text{ and } v=u)\\
\{(\hookleftarrow v)\} & \text{if }p(u)= 0 \text{ and } p(v)>0\\
\emptyset& \text{otherwise},
    \end{cases}
\end{equation*}
\begin{equation}
I_2=\{i:p_{i+1}\in R_1, p_{i+2}\in R_1, p_{i+1}\sim p_{i+2}\},
\end{equation}
and for $i\in I_2$ with $i$ read modulo $k$, let
\begin{equation}
D_i^2=(p_i\hookrightarrow p_{i+3})\backslash f_2(p_i,p_{i+3}).
\end{equation}
Then, the set $\{D_i^2:i\in I_2\}$ contains the largest feasible segment which has two articulation points.

\end{lemma}

\proof
We will first show that if $\{p_i,p_{i+1},p_{i+2},p_{i+3}\}\subset A(C)$ with $p_{i+1}\in R_1$, $p_{i+2}\in R_1$, and $p_{i+1}\sim p_{i+2}$, then $S:=(p_i\hookrightarrow p_{i+3})\backslash f(p_i,p_{i+3})$ is a feasible segment. First note that $S$ is indeed a segment, since $f_2(p_i,p_{i+3})$ can only remove the leaves of $G[(p_i\hookrightarrow p_{i+3})]$ from $(p_i\hookrightarrow p_{i+3})$. If $R:=M\cup C\backslash S$ is a set of initially colored vertices, any uncolored vertex in a pendant path, except the ones adjacent to $p_{i+1}$ and $p_{i+2}$, can be forced either by its base --- if its base is in $M$ --- or by the vertex the pendant path is attached to --- if its base is not in $M$. This includes any pendent paths attached to $p_i$ and $p_{i+3}$, since if they exist, $(p_i \hookrightarrow)$ and $(\hookleftarrow p_{i +3})$ would respectively be added to the forcing set by $f_2$, ensuring that the bases of these paths are the only uncolored neighbors of $p_i$ and  $p_{i+3}$. Thus, both ends of $S$ are either able to initiate a forcing chain reaching $p_{i+1}$ and $p_{i+2}$, or are themselves $p_{i+1}$ or $p_{i+2}$ (if $p_i$ happens to be adjacent to $p_{i+1}$, or if $p_{i+2}\sim p_{i+3}$). In either case, $p_{i+1}$ and $p_{i+2}$ will be colored at some step of the forcing process, whereupon each will be able to force their respective uncolored attached pendant paths. Thus, $R$ is a forcing set of $G$. 

We will now show that $S$ is maximal, by showing that if either end of $S$ is removed from $R$, the resulting set would not be forcing or would not be connected. First note that if $p(p_i)=0$ and $p(p_{i+3})=0$, $S$ is clearly maximal since by Lemma \ref{two_ap_lemma}, neither $p_i$ nor $p_{i+3}$ can be excluded from the forcing set. Next, note that $p_{i+1}$ and $p_{i+2}$ must be forced by two distinct forcing chains, since if a single forcing chain were to force them, then the first of $p_{i+1}$ and $p_{i+2}$ to be forced would have two uncolored neighbors, and could not force the other. Thus, if one or both of $p_i$ and $p_{i+3}$ are attached to a pendant path, then $(p_i\hookrightarrow )$ and $(\hookleftarrow p_{i+3})$ cannot be removed from $R$ since then one or both ends of the segment would not be able to initiate a forcing chain. 

In the special case of $p_i=p_{i+3}$, which happens when $|A(C)=3|$, $p_{i+1}$ and $p_{i+2}$ must still be forced by two distinct forcing chains; if $p_i$ is attached to a pendant path, then both its clockwise and counterclockwise neighbors must be added to the forcing set; the first case in the definition of $f_2$ remains valid for this situation. If $p_i$ is not attached to a pendant path, then one of its neighbors (say, the counterclockwise one) must nevertheless be added to the forcing set since $p_i$ cannot initiate two distinct forcing chains on its own. This is reflected in the second case of the definition of $f_2$.

Thus, every segment in $\mathcal{D}^2:=\{D_i^2:i\in I_2\}$ is feasible. Suppose there is a feasible segment $S'$ which contains two articulation points $p$ and $q$ but which is not in $\mathcal{D}^2$. The articulation points $p$ and $q$ must be adjacent, since otherwise $(p\hookrightarrow q)\neq \emptyset$ and by a similar argument as in Lemma \ref{two_ap_lemma}, the vertices in $(p\hookrightarrow q)$ cannot be forced. Moreover, by Lemma \ref{two_ap_lemma}, $p$ and $q$ must be attached only to single pendant paths; thus, they are some adjacent $p_{i+1}$ and $p_{i+2}$ in $R_1$; note that $p_i$ and $p_{i+3}$ exist (and are possibly equal), since by assumption $|A(C)|\geq 3$. By a similar argument as above, $S'$ cannot contain either end of $S:=(p_i\hookrightarrow p_{i+3})\backslash f_2(p_i,p_{i+3})$ so $S'$ can be at most equal to $S$. Moreover, since $S'$ is maximal, it cannot be a proper subset of $S$ since we have shown that $M\cup C\backslash S$ is a forcing set of $G$; therefore, $S'$ is precisely equal to $S$. Thus, by construction, $\mathcal{D}^2$ contains every feasible segment which has two articulation points, and in particular, the largest one. Note that $\mathcal{D}^2$ could also contain some segments that have fewer articulation points, which happens if $p_{i+1}$ or $p_{i+2}$ is subtracted from $(p_i\hookrightarrow p_{i+3})$ by $f_2(p_i,p_{i+3})$.
\qed

\begin{lemma}
\label{d1_lemma}
Let $G$ be a unicyclic graph, $C$ be the vertex set of the cycle of $G$ and suppose $|A(C)|\geq 2$. Let 
\begin{equation*}
f_1(u,v,w)=\begin{cases}
f_2(u,w) & \text{if }u\not\sim v\text{ and } v\not\sim w\\
\{(\hookleftarrow w)\}& \text{if }u\sim v \text{ and }v\not\sim w\text{ and }p(w)>0\\
\{(u\hookrightarrow)\}& \text{if }u\not\sim v \text{ and }v\sim w \text{ and }p(u)>0\\
v & \text{if }u\sim v, \,v\sim w,\, u\neq w,\,p(u)>0, \,p(w)>0\\
\{(u\hookrightarrow),(\hookleftarrow u)\}\backslash \{v\} & \text{if } u=w \text{ and } u\sim v\\
\emptyset& \text{otherwise},
    \end{cases}
\end{equation*}

\begin{equation}
I_1=\{i:p_{i+1}\in R_1\},
\end{equation}
and for $i\in I_1$ with $i$ read modulo $k$, let
\begin{equation}
D_i^1=(p_i\hookrightarrow p_{i+2})\backslash f_1(p_i,p_{i+1},p_{i+2}).
\end{equation}
Then, the set $\{D_i^1:i\in I_1\}$ contains the largest feasible segment which has one articulation point.
\end{lemma}
\proof

We will first show that if $\{p_i,p_{i+1},p_{i+2}\}\subset A$ with $p_{i+1}\in R_1$, then $S:=(p_i\hookrightarrow p_{i+2})\backslash f_1(p_i,p_{i+1},p_{i+2})$ is a maximal segment containing at most one articulation point for which $R:=M\cup C\backslash S$ is a forcing set of $G$. First, by a similar argument as in Lemma \ref{d2_lemma}, all pendant paths of $G$ attached to vertices other than $p_i$ and $p_{i+2}$ can get forced by their bases or the vertices to which they are attached. 
If neither $p_i$ nor $p_{i+2}$ is adjacent to $p_{i+1}$, then by a similar argument as in Lemma \ref{d2_lemma}, two separate forcing chains are needed to color $p_{i+1}$ and the pendant path attached to it; the first case in the definition of $f_1$ assures that this can happen in the same way as when the segment contains two articulation points, and that any pendant paths attached to $p_i$ and $p_{i+2}$ whose bases are not in $M$ get colored as well. If $p_i$ (but not $p_{i+2}$) is adjacent to $p_{i+1}$ and if $p_{i+2}$ is attached to a pendant path, then $f_1$ adds $(\hookleftarrow p_{i+2})$ to the forcing set, which initiates a forcing chain to color $S$ and allows $p_{i+1}$ to force its attached pendant path; then $p_i$ and $p_{i+2}$ will also be able to force any pendant paths attached to them whose bases are not in $M$.
Similarly, if $p_{i+2}$ is not attached to a pendant path, then it is able to initiate a forcing chain to color $S$ on its own; by symmetry, the same argument shows that $S$ gets colored if $p_{i+2}$ (but not $p_i$) is adjacent to $p_{i+1}$. If both $p_i$ and $p_{i+2}$ are adjacent to $p_{i+1}$, then $p_{i+1}$ must be added to the forcing set only if both $p_i$ and $p_{i+2}$ are attached to pendant paths; this is reflected in the fourth case of the definition of $f_1$.

Finally, in the special case of $p_i=p_{i+2}$, which happens when $C$ has 2 articulation points, there are several possible situations. If $p_i\not\sim p_{i+1}$, there must again be two forcing chains initiating outside $S$ which force $p_{i+1}$; the first line of the definition of $f_1$ is valid for this case, by a similar reasoning as in the special case of Lemma \ref{d2_lemma}. If $p_i\sim p_{i+1}$ and $p_i$ is attached to a pendant path, then a neighbor of $p_i$ in $C$ different from $p_{i+1}$ must be added to $R$ by $f_1$, so that this neighbor can initiate a forcing chain around $C$ to $p_{i+1}$ and the pendant path attached to it. If $p_i\sim p_{i+1}$ and $p_i$ is not attached to a pendant path, then any neighbor of $p_i$ in $C$ can be added to $R$ by $f_1$ to ensure $G$ is forced (including the one different from $p_{i+1}$). This is reflected in the fifth case of the definition of $f_1$. Thus, we have seen that in all cases, $R$ is a forcing set of $G$.

We will now show that $S$ is maximal, by showing that if either end of $S$ is removed from $R$, the resulting set would not be forcing, or would not be connected, or would contain two articulation points. First note that if neither  $p_i$ nor $p_{i+2}$ is adjacent to $p_{i+1}$, then by the same reasoning as in Lemma \ref{d2_lemma}, $S$ is maximal. If $p_i\sim p_{i+1}$, then the other end of $S$ must be able to initiate a forcing chain. Thus, if $p_{i+2}$ is attached to a pendant path, then $(\hookleftarrow p_{i+2})$ cannot be removed from $R$ since then $p_{i+2}$ would not be able to initiate a forcing chain. Similarly, if $p_{i+2}\sim p_{i+1}$, $(p_i \hookrightarrow)$ cannot be removed.

Thus, every segment in $\mathcal{D}^1:=\{D_i^1:i\in I_1\}$ is 
a maximal segment containing at most one articulation point, whose exclusion from $M\cup C$ yields a forcing set of $G$. Suppose there is a feasible segment $S'$ containing one articulation point $p$ which is not in $\mathcal{D}^1$. By Lemma \ref{two_ap_lemma}, the articulation point $p$ must be attached only to a single pendant path. Thus, this is some $p_{i+1}$ in $R_1$; note that $p_i$ and $p_{i+2}$ exist (and are possibly equal), since by assumption $|A(C)|\geq 2$. By a similar argument as above, $S'$ cannot contain either end of $S:=(p_i\hookrightarrow p_{i+2})\backslash f_1(p_i,p_{i+1},p_{i+2})$, up to the arbitrary choice made by $f_1$ when it must subtract one of two possible vertices from the segment in order to assure the resulting set is forcing, which does not affect the size of the segment. Thus, $S'$ can be at most equal in size to $S$; moreover, since $S'$ is maximal, it cannot be a proper subset of $S$ since we have shown that $M\cup C\backslash S$ is a forcing set of $G$. Thus, by construction, $\mathcal{D}^1$ contains a feasible segment of maximum size among all feasible segments with one articulation point.
\qed

\begin{lemma}
\label{d0_lemma}
Let $G$ be a unicyclic graph, $C$ be the vertex set of the cycle of $G$ and suppose $|A(C)|\geq 1$. Let 
\begin{equation*}
f_0(u,v)=\begin{cases}
(u\hookrightarrow) & \text{if }(p(u)\geq 1 \text{ and } p(v)\geq 1) \text{ or } u=v\\
\emptyset& \text{otherwise},
    \end{cases}
\end{equation*}

\begin{equation}
I_0=\{1,\ldots,k\},
\end{equation}
and for $i\in I_0$ with $i$ read modulo $k$, let
\begin{equation}
D_i^0=(p_i\hookrightarrow p_{i+1})\backslash f_0(p_i,p_{i+1}).
\end{equation}
Then, the set $\{D_i^0:i\in I_0\}$ contains the largest  feasible segment which has no articulation points.
\end{lemma}

\proof
We will first show that if $\{p_i,p_{i+1}\}\subset A(C)$, $S:=(p_i\hookrightarrow p_{i+1})\backslash f_0(p_i,p_{i+1})$ 
is a maximal segment containing no articulation points for which $R:=M\cup C\backslash S$ is a forcing set of $G$. First, by a similar argument as in Lemma \ref{d2_lemma}, all pendant paths of $G$ --- except any pendant paths attached to $p_i$ and $p_{i+1}$ whose bases are not in $M$ --- can get forced by their bases or by the vertices to which they are attached. If at least one of $p_i$ and $p_{i+1}$ is not attached to a pendant path, then the vertex not attached to a pendant path can initiate a forcing chain which colors $(p_i\hookrightarrow p_{i+1})$, and then the other vertex will be able to force its pendant path whose base is not in $M$, if it exists. Similarly, if both $p_i$ and $p_{i+1}$ are attached to pendant paths, then $f_0$ adds $(p_i\hookrightarrow)$ to the forcing set, which is able to initiate a forcing chain to color $S$. Note that if $p_i\sim p_{i+1}$, $R=M\cup C$ regardless of whether $p_i$ and $ p_{i+1}$ are attached to pendant paths. Thus, $R$ is a forcing set of $G$. $R$ is also maximal, since if either end of $S$ is removed from $R$, the resulting set would either contain an articulation point, or would not be forcing, or would not be connected. 

In the special case of $p_i=p_{i+1}$, which happens when $G$ has a single articulation point $p_1$, regardless of whether or not $p_1$ is attached to a pendant path, one of its neighbors in $C$ --- say, the counterclockwise one --- must be added to $R$ in order to initiate a forcing chain. This is reflected in the first case of the definition of $f_0$. 

Thus, every segment of $\mathcal{D}^0:=\{D_i^0:i\in I_0\}$ is a maximal segment containing no articulation points whose exclusion from $M\cup C$ yields a forcing set of $G$. Moreover, every feasible segment containing no articulation points must either have two ends which are articulation points, at least one of which is not attached to a pendant path, or have one end which is an articulation point attached to a pendant path, and another end which is a neighbor of an articulation point attached to a pendant path --- otherwise the segment would contain an articulation point, or would not be forcing, or would not be maximal. Thus, by construction, $\mathcal{D}^0$ contains every feasible segment which has no articulation points, up to the arbitrary choice of whether $f_0(u,v)$ subtracts $(u\hookrightarrow)$ or $(\hookleftarrow v)$, which does not affect the size of the segment. In particular, $\mathcal{D}^0$ contains a feasible segment of maximum size among all feasible segments with no articulation points.
\qed
\vspace{9pt}
For an illustration of the constructions in Lemmas \ref{d2_lemma}, \ref{d1_lemma}, and \ref{d0_lemma}, see Figure \ref{fig_unicycle} which shows a unicyclic graph with feasible segments of maximum size containing zero, one, and two articulation points.

\begin{theorem}
\label{th_unicyclic}
Let $G$ be a unicyclic graph and $C$ be the vertex set of the cycle of $G$. For $0\leq j\leq 2$, if $|A(C)|> j$, let $D^j_{max}=\max_{i\in I_j}\{|D^j_i|\}$, where $I_j$ and $D^j_i$ are defined as in (1)---(6); if $|A(C)|\leq j$, let $D^j_{max}=0$. Let $i^*$ and $j^*$ be such that $|D^{j^*}_{i^*}|=\max\{D^0_{max},D^1_{max},D^2_{max}\}$. Then,

\begin{equation*}
Z_c(G)=\begin{cases}
      2 & \text{if }|A(C)|=0\\
|M\cup C\backslash D^{j^*}_{i^*}| & \text{if }|A(C)|\geq 1,
    \end{cases}
\end{equation*}
and a minimum connected forcing set of $G$ can be found in $O(n)$ time.
\end{theorem}
\proof
If $|A(C)|=0$, then $G$ is a cycle, and two adjacent vertices of $G$ clearly form a minimum connected forcing set. Thus, we will henceforth assume $|A(C)|\geq 1$. By Lemma \ref{two_ap_lemma}, a feasible segment can have at most two articulation points. By Lemmas \ref{d2_lemma}, \ref{d1_lemma}, and \ref{d0_lemma}, $D^{j^*}_{i^*}$ is the largest feasible segment of $C$, and by Lemma \ref{rstarlemma}, $|M\cup C\backslash D^{j^*}_{i^*}|$ is a minimum connected forcing set of $G$.

To verify that the time needed to find $D^{j^*}_{i^*}$ is linear in the order of the graph, first note that the set of articulation points in $G$, and hence the points in $M$, $C$, and $A(C)$, can be found in linear time (cf. \cite{tarjan}). Then, the sets $(p_i\hookrightarrow p_{i+1})$, $1\leq i\leq k$, can also be found in linear time. These sets of articulation points and vertices can be stored (with linear space), and each of the functions $f_0$, $f_1$, and $f_2$ and $D_i^j$ can be computed in constant time for 0$\leq j\leq 2$ and $1\leq i\leq k$. Since each of the index sets $I_0$, $I_1$, and $I_2$ has at most $|A(C)|=O(n)$ elements, $D^{j^*}_{i^*}$ can be found by computing the maximum of $O(n)$ terms.
\qed

\vspace{9pt}
The zero forcing number and path cover number of unicyclic graphs have been investigated in \cite{fallat,row,taklimi} and have been shown to coincide. We conclude this section by characterizing the unicyclic graphs for which $Z(G)=Z_c(G)$, and thus describing the connectivity of the minimum zero forcing sets of unicyclic graphs.

\begin{figure}[h!]
\begin{center}
\includegraphics[scale=0.35]{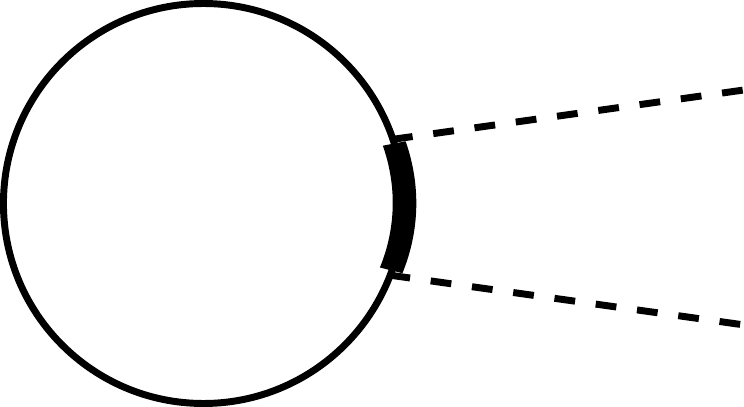}
\caption{Unicyclic graphs for which the zero forcing number equals the connected forcing number. The solid line indicates a cycle of arbitrary size; the bold line indicates a single edge; the dotted lines indicate paths of arbitrary (possibly zero) length.}
\label{fig_eq}
\end{center}
\end{figure}

\begin{proposition}
\label{prop_eq}
For a unicyclic graph $G$, $Z_c(G)=Z(G)$ if and only if $G$ is in the family of graphs depicted in Figure \ref{fig_eq}.
\end{proposition}

\proof
Let $G$ be a unicyclic graph satisfying $Z_c(G)=Z(G)$. Let $C$ be the vertex set of the cycle of $G$, let $R$ be an arbitrary minimum connected forcing set of $G$, and fix an arbitrary chronological list of forces for $R$. Suppose for contradiction that $G$ has a vertex $v\in C\cap(R_2\cup R_3)$. If there is a component $T$ of $G-v$ which is a tree different from a path, then the vertex $u$ in $T$ which is adjacent to $v$ (in $G$) is also in $R_2\cup R_3$, and is therefore not forced by $v$. Let $Z$ be the set obtained by removing $v$ from $R$ and replacing each vertex of $T$ in $R$ which is the base of a pendant path or which is a vertex in $R_3$ that forces a pendant path, by the leaf of that pendant path. In other words, $Z$ is obtained by reversing each forcing chain contained in $T$. We claim that $Z$ is a zero forcing set of $G$. To see why, note that each colored leaf will force its pendant path and the vertex to which it is attached; thus, whether or not $u$ forces a vertex in $T$, at some point in the forcing process, $u$ will be colored along with all its neighbors except $v$. Then, $u$ will force $v$, whereupon the set of colored vertices will include $R$ and will thus be able to color the rest of $G$. On the other hand, if no component of $G-v$ is a tree different from a path, then since $v\in R_2\cup R_3$, there must be at least two pendant paths attached to $v$. The set $Z$ obtained by removing $v$ from $R$ and replacing the base of one of the pendants which is in $M$ by the leaf of that pendant is also a zero forcing set of $G$ by a similar reasoning as in the case above. Thus, if $C\cap(R_2\cup R_3)\neq \emptyset$, then $Z(G)\leq |Z|<|R|=Z_c(G)$. It follows that $C\cap(R_2\cup R_3)=\emptyset$ and that each articulation point of $C$ is in $R_1$. 

Now, for any $v\in C$, define $\ell(v)$ to be $v$ if $v$ is not an articulation point, and to be the leaf of the pendant path attached to $v$ if $v$ is an articulation point. Suppose for contradiction that $|R\cap C|\geq 3$. As shown in Lemma \ref{one_seg_lemma}, $R$ can exclude at most one segment of $C$, which implies that $R\cap C$ also forms a segment; thus, there are vertices $\{u,v,w\}\subset R\cap C$ such that $u\sim v$ and $v\sim w$. We claim $Z:=R\backslash\{u,v,w\}\cup\{\ell(u),\ell(v)\}$ is a zero forcing set of $G$. To see why, note that whether or not $u$ and $v$ are attached to pendant paths, at the first stage of the forcing process, $\ell(u)$ and $\ell(v)$ can initiate forcing chains which color $u$ and $v$; next, $w$ will be the only uncolored neighbor of $v$, and $v$ can force $w$. At this point, the set of colored vertices contains $R$, and can therefore color all of $G$; this means $Z(G)\leq |Z|<|R|=Z_c(G)$ --- a contradiction. Thus, $|R\cap C|<3$; on the other hand, by Proposition \ref{block_prop}, $|R\cap C|\geq 2$, so $|R\cap C|=2$. By inspection, only the unicyclic graphs in Figure \ref{fig_eq} satisfy this condition and the condition that each articulation point of $C$ is in $R_1$. 
\qed

\begin{remark}
Note that Figure \ref{fig_eq} does not include all unicyclic graphs with zero forcing number 2; one can easily find a unicyclic graph $G$ with $Z(G)=2$ and $Z_c(G)>2$. 
\end{remark}

\noindent Proposition \ref{prop_eq} allows us to make the following characterization of the minimum zero forcing sets of unicyclic graphs.

\begin{corollary}
For all unicyclic graphs except the ones in Figure \ref{fig_eq}, any minimum zero forcing set is disconnected.
\end{corollary}

\subsection{Cactus and block graphs with no pendant paths}
We will now characterize the connected forcing numbers of cactus and block graphs which have no pendant paths. The general case is deferred to future work. Before we begin, we need the following definition.

\begin{definition}
Let $G$ be a graph, let $G_0=G$, and for $i\geq 1$, let $G_i=G_{i-1}-\{$all non-articulation points of the outer blocks of $G_{i-1}\}$. We will say a block of $G$ has \emph{depth} $i$ if it is an outer block of $G_i$.
\end{definition}

\begin{proposition}
\label{block_graph_prop}
Let $G=(V,E)$ be a block graph with no pendant paths and $b$ be the number of blocks of $G$ which have at least one non-articulation vertex. Then $Z_c(G)=n-b$.
\end{proposition}
\proof
Let $Q$ be a set containing one non-articulation vertex from each block of $G$ which has non-articulation vertices. We claim that $R:=V\backslash Q$ is a minimum connected forcing set of $G$. $G[R]$ is clearly connected, since deleting one non-articulation vertex from each block by definition does not disconnect $G$. Next, since $G$ has no pendant paths, each outer block of $G$ has size at least 3; thus, each outer block has at least two non-articulation vertices, one of which is in $Q$ and the other of which is in $R$ and can force the first. Thus, each block at depth 0 can be forced. Now suppose every block at depth at most $i\geq 0$ has been colored and let $B$ be a block at depth $i+1$. By definition, $B$ must have an articulation point $p$ adjacent only to blocks at depth less than $i+1$ (besides $B$), since otherwise $B$ would not be an outer block when all blocks of smaller depth are deleted. Since all blocks adjacent to $p$ besides $B$ have been colored by assumption, $p$ can force an uncolored non-articulation vertex in $B$, if such a vertex exists. Thus, each block at depth $i+1$ will get colored as well. By induction, every block in the graph can get forced by $R$, so $R$ is a connected forcing set. 

Now, let $S$ be an arbitrary minimum connected forcing set of $G$. By Proposition \ref{block_prop}, $S$ must contain at least $\delta(G[B])=|B|-1$ vertices from each block $B$ of $G$. Moreover, since $G$ has no pendant paths, all articulation points of $G$ are in $R_2$ or $R_3$, so by Lemma \ref{MR_lemma}, all articulation points of $G$ must be in $S$. Thus, $S$ can exclude at most one vertex from each of the $b$ blocks that have non-articulation vertices, so $|S|\geq n-b$. Thus, $R$ is a minimum connected forcing set.
\qed

\begin{proposition}
\label{cactus_graph_prop}
Let $G=(V,E)$ be a cactus graph with no pendant paths. Let $\mathcal{C}$ be the collection of vertex sets of cycles of $G$ and $b$ be the number of outer blocks of $G$. For $C\in \mathcal{C}$, let $D_C$ be the largest segment of $C$ which does not contain articulation points of $C$. Then, $Z_c(G)=n-\sum_{C\in \mathcal{C}}|D_C|+b$, if $G$ is not a cycle, and $Z_c(G)=2$ if $G$ is a cycle.
\end{proposition}
\proof
Clearly $Z_c(G)=2$ if $G$ is a cycle, so suppose henceforth that $G$ is not a cycle. Let $Q$ be a set containing one vertex from each outer block of $G$ which is adjacent to the articulation point of the outer block. Let $\mathcal{D}:=\bigcup_{C\in \mathcal{C}}D_C$ and $R:=(V\backslash \mathcal{D})\cup Q$. We claim that $R$ is a minimum connected forcing set of $G$. $R$ is connected, since deleting one segment containing no articulation points from each cycle block does not disconnect $G$. Next, since $G$ has no pendant paths, each outer block of $G$ is a cycle in which the articulation point and one of its neighbors (the one in $Q$) are in $R$. In each outer block, the colored neighbor of the articulation point will initiate a forcing chain around the cycle; thus, each block at depth 0 can be forced. Now suppose every block at depth at most $i\geq 0$ has been colored and let $B$ be a block at depth $i+1$. If $B$ is a cut edge block, then both vertices of $B$ are already in $R$. If $B$ is a cycle block, then let $(u\hookrightarrow v)$ be the segment missing from $B$. By definition, one of $u$ and $v$ --- say, $u$ --- must be adjacent only to blocks at depth less than $i+1$ (besides $B$), since otherwise $B$ would not be an outer block when all blocks of smaller depth are deleted. Since all blocks adjacent to $u$ besides $B$ have been colored by assumption, $u$ can initiate a forcing chain which colors the segment $(u\hookrightarrow v)$. Thus, each block at depth $i+1$ will get colored as well. By induction, every block in the graph can get forced by $R$, so $R$ is a connected forcing set. 

Finally, suppose there is a connected forcing set $R'$ with $|R'|<|R|$. Let $B$ be the vertex set of some block of $G$ which contains fewer vertices of $R'$ than of $R$. $B$ cannot be the vertex set of a cut edge block of $G$, since both vertices in each cut edge block are in $R_2$ and hence belong to $R'$ by Lemma \ref{MR_lemma}. $B$ also cannot be an outer block, since every outer block is a cycle and by Proposition~\ref{block_prop}, $R'$ contains at least $\delta(G[B])=2$ vertices from each such block. Thus, $B$ must be a non-outer cycle block of $G$. However, by Lemma \ref{one_seg_lemma}, the vertices of $B$ which $R'$ excludes form a segment of $B$, but by construction, this segment cannot be bigger than the segment excluded from $R$. This is a contradiction, so $R$ is a minimum connected forcing set. Since the segments $\{D_C:C\in \mathcal{C}\}$ are disjoint, $Z_c(G)=n-\sum_{C\in \mathcal{C}}|D_C|+b$.
\qed

\begin{remark}
The characterizations of Propositions \ref{block_graph_prop} and \ref{cactus_graph_prop} are constructive, and minimum connected forcing sets of cactus and block graphs with no pendant paths can be found in linear time, by a similar analysis as in Theorem \ref{th_unicyclic}.
\end{remark}

\section{Connected forcing and matroids}

In this section, we investigate a relation between connected forcing sets, greedoids, and matroids. A \emph{matroid} is an ordered pair $(S,\mathcal{I})$ where $S$ is a finite set and $\mathcal{I}$ is a subset of $\mathcal{P}(S)$ (the power set of $S$) satisfying 
\begin{enumerate}
\item[](M1) $\emptyset\in \mathcal{I}$
\item[](M2) If $J'\subset J\in \mathcal{I}$ then $J'\in \mathcal{I}$
\item[](M3) For every $A\subset S$, every maximal subset of $A$ in $\mathcal{I}$ has the same cardinality.
\end{enumerate}

\noindent An ordered pair $(S,\mathcal{I})$ which satisfies only (M1) and (M3) is called a \emph{greedoid}.
Matroids and greedoids have been studied extensively; see, e.g., \cite{korte_lovasz,oxley} for some of their fundamental properties, and in particular their connection to the greedy algorithm. We can define a \emph{greedy algorithm} for finding a connected forcing set of a graph $G=(V,E)$ as follows:

\begin{itemize}
\item[]Set $R=V$;
\item[]While there exists $v\in R$ with $R\backslash \{v\}$ being a connected forcing set,
\item[]\qquad Replace $R$ by $R\backslash \{v\}$.
\end{itemize}

\noindent Clearly, this algorithm always produces a connected forcing set. Our next results show that in some graphs, the greedy algorithm produces a minimum connected forcing set, and that the collection of all connected forcing sets can be used to define greedoids and matroids.

\begin{theorem}
\label{greedoid_matroid}
Let $\mathcal{T}$ be the family of trees, $\mathcal{T}'$ be the family of trees whose pendant paths have length one, and
$\mathcal{B}$ be the family of block graphs with no pendant paths. 
\begin{enumerate}
\item Let $G=(V,E)\in \mathcal{T}\cup\mathcal{B}$, $G\not\simeq P_n$, and let $\mathcal{I}$ be the set of all connected forcing sets of $G$. Then $(V,\mathcal{P}(V)\backslash\mathcal{I})$ is a greedoid.
\item Let $G=(V,E)\in \mathcal{T}'\cup\mathcal{B}$, $G\not\simeq P_n$, and let $\mathcal{I}$ be the set of all connected forcing sets of $G$. Then $(V,\mathcal{P}(V)\backslash\mathcal{I})$ is a matroid.
\end{enumerate}
\end{theorem}
\proof
Suppose $G$ is a tree different from a path and let $A\subset V$. If $a\in A\backslash (R_2\cup R_3)$, then $a$ belongs to a pendant path of $G$. Let $X_1(A)$ be the set containing, for all $a\in A\backslash (R_2\cup R_3)$, the vertices of the pendant path containing $a$ which lie between $a$ and the base of that pendant path, including $a$ and the base of the path. Let $X_2(A)$ be the set containing, for $v\in R_3$, all-but-one bases of pendant paths attached to $v$ which do not belong to pendant paths containing vertices of $A$. We claim that a minimal superset $S$ of $A$ which is a connected forcing set of $G$ is the union of $R_2$, $R_3$, $X_1(A)$ and $X_2(A)$. First, note that $S$ is clearly a superset of $A$ since $A\subset R_2\cup R_3\cup X_1(A)$; $S$ is also connected, since the only vertices of $G$ which are not in $S$ are connected parts of some pendant paths which contain the leaves of those pendant paths, and deleting those does not disconnect $G$. $S$ is also forcing, since it contains $M$, which by Theorem \ref{tree_thm} is a minimum connected forcing set of $G$. Now suppose for contradiction that for some $s\in S$, $S\backslash \{s\}$ is also a connected forcing set. By Lemma \ref{MR_lemma}, $R_2\cup R_3\subset S$, so $s\notin R_2\cup R_3$. Since each $a\in A\backslash R_2\cup R_3$ is in $S$, since the vertex attached to the pendant path containing $a$ is in $S$, and since $S$ is connected, $S$ must also contain all vertices in that pendant path which lie between $a$ and the base of the pendant path; thus $s\notin X_1(A)$. Finally, by Lemma \ref{MR_lemma}, $S$ must contain all-but-one bases of pendant paths attached to each $v\in V$; thus, $s\notin X_2(A)$. Therefore, $S$ is minimal. Since $R_2$, $R_3$ and $X_1(A)$ are determined by the structure of $G$ and the given set $A$, and the arbitrary choice of bases in $X_2(A)$ does not affect the cardinality of $X_2(A)$, every minimal superset $S$ of $A$ which is a connected forcing set of $G$ has the same cardinality.

Next, suppose $G$ is a block graph which has no pendant paths and is different from a path and let $A\subset V$. Let $X(A)$ be the set containing, for each block $B$ of $G$, one non-articulation vertex in $B$ which is not in $A$, if such a vertex exists. We claim that a minimal superset $S$ of $A$ which is a connected forcing set of $G$ equals $V\backslash X(A)$. First, note that by construction, $S$ is a superset of $A$. $S$ is also connected, since it excludes only non-articulation points of $G$, and $S$ is forcing, since it contains a minimum connected forcing set of $G$, namely, $R$ as defined in the proof of Proposition \ref{block_graph_prop}. Suppose there is some $s\in S$ such that $S\backslash\{s\}$ is also a connected forcing set of $G$. By Lemma \ref{MR_lemma}, $s$ is a non-articulation vertex of some block $B$, and $s\notin A$. However, since $X(A)$ contains a vertex from each block which has a non-articulation point which is not in $A$, $X(A)$ must already include a vertex from $B$. However, by Proposition \ref{block_prop}, $S$ cannot exclude two vertices from $B$. Thus $S$ is minimal, and since the arbitrary choice of vertices in $X(A)$ does not affect the cardinality of $X(A)$, every minimal superset $S$ of $A$ which is a connected forcing set of $G$ has the same cardinality.

For any $G=(V,E)\in \mathcal{T}\cup \mathcal{B}$, $V$ is clearly a connected forcing set of $G$. Thus, the ordered pair $(V,\mathcal{I})$ satisfies

\begin{enumerate}
\item[](M$1'$) $V\in \mathcal{I}$
\item[](M$3'$) For every $A\subset V$, every minimal superset of $A$ in $\mathcal{I}$ has the same cardinality.
\end{enumerate}

\noindent Now, it can be verified that $(V,\mathcal{P}(V)\backslash \mathcal{I})$ satisfies properties (M1) and (M3) and is therefore a greedoid.

Suppose $G$ is a tree whose pendant paths have length one, and let $J$ be an arbitrary connected forcing set of $G$. Since by Lemma \ref{MR_lemma}, $M\subset J$, the only vertices of $G$ not in $J$ are some of the leaves of $G$. Let $J'$ be a superset of $J$. Since each leaf of $G$ is adjacent to a vertex in $J$ and since $J\subset J'$ is a forcing set of $G$, $J'$ is also a connected forcing set of $G$.

Suppose $G$ is a block graph with no pendant paths and let $J$ be an arbitrary connected forcing set of $G$. By Proposition \ref{block_graph_prop}, the only vertices of $G$ not in $J$ are up to one non-articulation vertex in each block of $G$. Let $J'$ be a superset of $J$. Since each non-articulation vertex of $G$ is adjacent to a vertex in $J$ and since $J\subset J'$ is a forcing set of $G$, $J'$ is also a connected forcing set of $G$.

Thus, for any $G=(V,E)\in \mathcal{T}'\cup \mathcal{B}$, the ordered pair $(V,\mathcal{I})$ satisfies
\begin{enumerate}
\item[](M$2'$) If $J'\supset J\in \mathcal{I}$ then $J'\in \mathcal{I}$.
\end{enumerate}
Moreover, since $\mathcal{T}'\cup \mathcal{B}\subset \mathcal{T}\cup \mathcal{B}$, $(V,\mathcal{I})$ satisfies properties (M1$')$ and (M$3')$ as well. Thus, it can be verified that $(V,\mathcal{P}(V)\backslash \mathcal{I})$ satisfies properties (M1), (M2), and (M3) and is therefore a matroid.
\qed
\vspace{9pt}

\noindent We will now briefly address several (negative) results related to Theorem \ref{greedoid_matroid}.
\begin{enumerate}
\item If $G=(V,E)$ is an arbitrary cactus or block graph (or a cactus graph with no pendant paths) and $\mathcal{I}$ is the collection of connected forcing sets of $G$, then $(V,\mathcal{P}(V)\backslash\mathcal{I})$ is not necessarily a greedoid. As a simple counterexample, let $G$ be the graph obtained by attaching two pendants to a triangle, each to a different vertex. Let $S_1$ be the set containing both vertices of $G$ of degree 3, and one vertex of degree 1, and $S_2$ be the set containing one vertex of degree 3 and one vertex of degree 2. $S_1$ and $S_2$ are minimal connected forcing sets, but do not have the same cardinality.
\item If $G=(V,E)$ is an arbitrary tree and $\mathcal{I}$ is the collection of connected forcing sets of $G$, then $(V,\mathcal{P}(V)\backslash\mathcal{I})$ is not necessarily a matroid, since a superset of a connected forcing set of $G$ could be disconnected.
\item If $G=(V,E)$ is a graph in $\mathcal{T}$, $\mathcal{T}'$, or $\mathcal{B}$ (defined as in Theorem \ref{greedoid_matroid}) and $\mathcal{I}$ is the collection of zero forcing sets of $G$, then $(V,\mathcal{P}(V)\backslash\mathcal{I})$ is not necessarily a greedoid or a matroid, since, for example, not every minimal zero forcing set of a graph in these families is minimum. This is only true for restricted subfamilies like star graphs, complete graphs, and cycles. In general, when defining matroids as in Theorem \ref{greedoid_matroid}, it appears that axiom (M$2)'$ is harder to satisfy for the collection of connected forcing sets, since a superset of a connected forcing is always forcing but not always connected, and (M$3)'$ is harder to satisfy for the collection of zero forcing sets, since there are no vertices which are part of every minimum zero forcing set of a graph.
\end{enumerate}

Even if the collection of connected forcing sets of a graph does not define a greedoid or a matroid, the greedy algorithm may nevertheless produce a minimum connected forcing set. We give an example of a family of graphs for which this is the case.

\begin{proposition}
Let $G$ be a cactus graph different from a path, all of whose cycles are outer blocks. Then, the greedy algorithm produces a minimum connected forcing set of $G$.
\end{proposition}
\proof
Let $Q$ be defined as in the proof of Proposition \ref{cactus_graph_prop}; by a similar argument as in Proposition \ref{cactus_graph_prop}, $M\cup Q$ is a minimum connected forcing set of $G$. Let $S$ be a minimal connected forcing set of $G$. By Lemma \ref{MR_lemma}, $M\subset S$, and by Proposition \ref{block_prop}, $S$ contains at least two vertices of each cycle. Since the articulation point of each cycle is in $S$ and $S$ is connected, at least one neighbor of the articulation point of each cycle must be in $S$. However, a single colored neighbor of the articulation point of each cycle is sufficient to initiate a forcing chain around the cycle; thus, for each cycle of $G$, $S$ contains exactly one neighbor of the articulation point of the cycle. 
Moreover, if $S$ contains a vertex $v$ which does not belong to $M$ or to any cycle of $G$, then $v$ must belong to a pendant path of $G$; however, $v$ and all other vertices from that pendant path (except one which is in $M$) can be removed from $S$, and the resulting set is still connected and forcing. Thus, $S$ does not contain any vertices outside $M\cup Q$. Therefore, every minimal connected forcing set of $G$ is also minimum. By definition, the greedy algorithm produces a minimal connected forcing set of $G$; thus, in this case it also produces a minimum connected forcing set.
\qed

\begin{remark}
The greedy algorithm has run time $O(n\cdot F(n))$, where $F(n)$ is the time required for checking whether a vertex set of size $n$ is forcing. In general, such an approach would take superlinear time, whereas the constructions for finding minimum connected forcing sets of trees and the graphs described in Propositions \ref{block_graph_prop} and \ref{cactus_graph_prop} can be realized in linear time. 
\end{remark}

\section{Conclusion}
In this paper, we have furthered the study of the connected variant of zero forcing. Connected forcing sets can be potentially applicable to modeling various physical phenomena, and the connected forcing number is a sharp upper bound to important graph parameters like the zero forcing number, path cover number, and maximum nullity. We have identified several structural results of connected forcing relating to forcing spread, graph density, and induced subgraphs, some of which can also be used in the study of zero forcing. We have also shown that the problem CZF is NP-complete. This result motivates the following question:

\begin{question}
\label{question1}
For which families of graphs is CZF solvable in polynomial time?
\end{question}

We have provided a partial answer to Question \ref{question1} by finding linear time algorithms for CZF in unicyclic graphs, as well as block graphs and cactus graphs with no pendant paths. Our second question relating to these findings is as follows.
\begin{question}
\label{question2}
What is the connected forcing number of cactus graphs and block graphs which have pendant paths?
\end{question}
More generally, it would be useful to develop a framework for computing the connected forcing number of a graph with cut vertices in terms of the connected forcing numbers of its blocks. Such a  framework has been developed for the zero forcing number (see, e.g., \cite{row}), but the same approach does not carry over to connected forcing due to the unboundedness of the connected forcing spread of vertices and edges.

Finally, we showed that for some families of graphs, the greedy algorithm produces minimum connected forcing sets, and the collections of connected forcing sets can be used to define greedoids and matroids. This motivates our last question, which is also a special case of Question \ref{question1}.
\begin{question}
For which families of graphs are CZF and ZF solvable by the greedy algorithm?
\end{question}

\section*{Acknowledgements}
This work is supported by the National Science Foundation, Grant No. 1450681.


\begin{thebibliography}{99}

\bibitem{aazami}
A. Aazami.
Hardness results and approximation algorithms for some
problems on graphs.
PhD thesis, University of Waterloo, 2008.

\bibitem{AIM-Workshop}
AIM Special Work Group.
Zero forcing sets and the minimum rank of graphs.
\emph{Linear Algebra and its Applications}, 428(7): 1628--1648, 2008.

\bibitem{zf_tw}
F. Barioli,  W. Barrett, S.M. Fallat, T. Hall, L. Hogben, B. Shader, P. van den Driessche, and H. van der Holst.
Parameters related to tree-width, zero forcing, and maximum nullity of a graph.
\emph{Journal of Graph Theory}, 72(2): 
146--177, 2013.


\bibitem{Barioli}
F. Barioli, W. Barrett, S. Fallat, H. T. Hall, L. Hogben, B. Shader, P. van den Driessche, and 
H. van der Holst. 
Zero forcing parameters and minimum rank problems. 
\emph{Linear Algebra and its Applications}, 433(2): 
401--411, 2010




\bibitem{benson}
K. Benson, D. Ferrero, M. Flagg, V. Furst, L. Hogben, V. Vasilevska, and B. Wissman.
Power domination and zero forcing.
\emph{arXiv}:1510.02421, 2015


\bibitem{pathcover1}
F. T. Boesch, S. Chen, and J. A. M. McHugh. 
On covering the points of a graph with point disjoint paths. 
\emph{Graphs and combinatorics}. Springer Berlin Heidelberg, 
201--212, 1974




\bibitem{bondy}
J. A. Bondy and U. S. R. Murty. 
\emph{Graph Theory with Applications}. Vol. 290. 
London, Macmillan, 1976

\bibitem{brimkov}
B. Brimkov and R. Davila. 
Characterizations of the connected forcing number of a graph.
\emph{arXiv}:1604.00740, 2016


\bibitem{quantum1}
D. Burgarth and V. Giovannetti.
Full control by locally induced relaxation.
\emph{Physical Review Letters}, 99(10): 
100501, 2007


\bibitem{logic1}
D. Burgarth, V. Giovannetti,  L. Hogben, S. Severini,  and M. Young.
Logic circuits from zero forcing.
\emph{arXiv}:1106.4403, 2011

\bibitem{Caro}
Y. Caro, D. B. West, and R. Yuster. 
Connected domination and spanning trees with many leaves
\emph{SIAM J. Discrete Math.}, 13: 
202--211, 2000


\bibitem{CF_paper}
R. Davila, M. Henning, C. Magnant, and R. Pepper. 
Bounds on the connected forcing number of a graph. \emph{arXiv}:1605.02124, 2016

\bibitem{Desormeaux}
W. J. Desormeaux, T. W. Haynes, and M. A. Henning. 
Bounds on the connected domination number of a graph.
\emph{Discrete Applied Mathematics}, 161(18): 
2925--2931, 2013

\bibitem{Edholm}
C. Edholm, L. Hogben, J. LaGrange, and D. Row.
Vertex and edge spread of zero forcing number, maximum nullity, and minimum rank of a graph.
\emph{Linear Algebra and its Applications}, 436(12): 4352--4372, 2012

\bibitem{positive_zf2}
J. Ekstrand, et al. 
Positive semidefinite zero forcing. 
\emph{Linear Algebra and its Applications}, 439(7):
1862--1874, 2013

\bibitem{Eroh}
L. Eroh, C. Kang, and E. Yi.
Metric dimension and zero forcing number of two families of line graphs.
\emph{arXiv}:1207.6127, 2012

\bibitem{fallat}
S. Fallat and L. Hogben.
 The minimum rank of symmetric matrices described by a graph: A survey. 
\emph{Linear Algebra and its Applications}, 426: 
558--582, 2007


\bibitem{connected_dom_1}
F. V. Fomin, F. Grandoni, and D. Kratsch. 
Solving connected dominating set faster than $2^n$. \emph{Algorithmica}, 52(2): 
153--166, 2008




\bibitem{GJ}
M. Garey and D. Johnson, 
\emph{Computers and Intractability}, 
W.H. Freeman \& Co., San Francisco, 1979


\bibitem{signed_zf}
F. Goldberg and A. Berman. 
Zero forcing for sign patterns. 
\emph{Linear Algebra and its Applications}, 447: 
56--67, 2014

\bibitem{pathcover2}
S. Goodman and S. Hedetniemi. 
On the Hamiltonian completion problem.
\emph{Graphs and combinatorics}. Springer Berlin Heidelberg, 
262--272, 1974




\bibitem{powerdom3}
T. Haynes,  S. Hedetniemi, S. Hedetniemi, and M. Henning.
Domination in graphs applied to electric power networks.
\emph{SIAM Journal on Discrete Mathematics}, 15(4): 
519--529, 2002


\bibitem{tarjan}
J. Hopcroft and R. Tarjan.
Algorithm 447: efficient algorithms for graph manipulation.
\emph{Communications of ACM} 16(6):
372--378, 1973

\bibitem{Huang}
L.-H. Huang, G. J. Chang, and H.-G. Yeh. 
On minimum rank and zero forcing sets of a graph.
\emph{Linear Algebra and its Applications}, 432: 2961--2973, 2010

\bibitem{johnson}
C. R. Johnson and A. Leal Duarte. 
The maximum multiplicity of an eigenvalue in a matrix
whose graph is a tree. 
\emph{Linear and Multilinear Algebra}, 46: 
139--144, 1999

\bibitem{korte_lovasz}
B. Korte and L. Lovasz. 
Mathematical structures underlying greedy algorithms. \emph{International Conference on Fundamentals of Computation Theory}. Springer Berlin Heidelberg, 1981

\bibitem{meeks}
K. Meeks and A. Scott. 
Spanning trees and the complexity of flood-filling games. \emph{Theory of Computing Systems}, 54(4): 
731--753, 2014

\bibitem{Meyer}
S. Meyer.
Zero forcing sets and bipartite circulants.
\emph{Linear Algebra and its Applications}, 436(4): 888--900, 2012


\bibitem{oxley}
J. G. Oxley. 
\emph{Matroid theory}. Vol. 3. Oxford University Press, USA, 2006


\bibitem{row}
D. D. Row. 
A technique for computing the zero forcing number of a graph with a cut-vertex.
\emph{Linear Algebra and its Applications}, 436: 4423--4432, 2012

\bibitem{Sampathkumar}
E. Sampathkumar and H. B. Walikar. 
The connected domination of a graph.
\emph{Math. Phys. Sci.}, 13: 
607--613, 1979

\bibitem{taklimi}
F. A. Taklimi. 
Zero forcing sets for graphs. 
\emph{arXiv}:1311.7672, 2013


\bibitem{zf_np}
M. Trefois and J. C. Delvenne.
Zero forcing number, constrained matchings and strong structural controllability.
\emph{arXiv}:1405.6222v2, 2015

\bibitem{power_dom_block}
G. Xu, L. Kang, E. Shan, and M. Zhao.
Power domination in block graphs. 
\emph{Theoretical Computer Science}, 359(1): 
299--305, 2006

\bibitem{powerdom2}
M. Zhao,  L. Kang, and G. Chang.
Power domination in graphs.
\emph{Discrete Mathematics}, 306(15): 
1812--1816, 2006






\end{thebibliography}
\end{document}